\documentstyle[prl,eqsecnum,aps,multicol,epsfig]{revtex}

\begin{document}
\draft
\title{
Low-Temperature Structure of the Quarter-Filled Ladder Compound $\alpha'$-NaV$_2$O$_5$
}
\author{Hiroshi Sawa,$^{1,}$\cite{byline} Emi Ninomiya,$^1$ Tetsuo Ohama,$^2$  Hironori Nakao,$^3$  Kenji Ohwada,$^4$  Youichi Murakami,$^{3,}$\cite{byline2} Yasuhiko Fujii,$^4$ Yukio Noda,$^5$ Masahiko Isobe,$^4$ and  Yutaka Ueda$^4$}
\address{
1 Graduate School of Science and Technology, Chiba University, Chiba 263-8522, Japan\\
2 Faculty of Science, Chiba University, Chiba 263-8522, Japan\\
3 Photon Factory, Institute of Materials Structure Science, KEK, Tsukuba 305-0801, Japan\\
4 Institute for Solid State Physics, The University of Tokyo, Kashiwa 277-8581, Japan\\
5 Institute of Multidisciplinary Research for Advanced Materials,Tohoku University, Sendai 980-8577, Japan\\
}

\date{\today}
\maketitle
\begin{abstract}
The low-temperature (LT) superstructure of $\alpha'$-NaV$_2$O$_5$ was determined by synchrotron radiation x-ray diffraction.
Below the phase transition temperature associated with atomic displacement and charge ordering at 34K, we observed the Bragg peak splittings, which evidence that the LT structure is monoclinic.
It was determined that the LT structure is $(a-b)\times 2b \times 4c$ with the space group $A112$ where $a, b$ and $c$ represent the high temperature orthorhombic unit cell.
The valence estimation of V ions according to the bond valence sum method shows that the V sites are clearly separated into two groups of V$^{4+}$ and V$^{5+}$ with a $zigzag$ charge ordering pattern.
This LT structure is consistent with resonant x-ray and NMR measurements, and  strikingly contrasts to the LT structure previously reported, which includes V$^{4.5+}$ sites.
\\
\end{abstract}

\begin{multicols}{2}
The low-dimensional mixed-valence vanadate $\alpha'$-NaV$_2$O$_5$ is known to show a novel phase transition at $T_{\rm C}$=34K\cite{Isobe}.
This transition is a charge ordering that involves a valence change of V ions as 2V$^{4.5+}\rightarrow$V$^{4+}+$V$^{5+}$\cite{Ohama1} but is an insulator-insulator transition \cite{Isobe3}, indicating that it does not originate in the Fermi surface instability.
In the high-temperature (HT) phase, V ions form quarter-filled two-leg ladders in layered trellis lattices(V$_2$O$_5$ planes).
The crystal structure is orthorhombic with the space group $Pmmn$ ($a_p$, $b_p$, $c_p$) at room temperature\cite{Meetsma,Schner,Smolin,Tsuda}.
The electronic Coulomb repulsion and strongly anisotropic electron hopping make the system an insulator in spite of the quarter-filling\cite{Smolin,Horsch}.
In the low-temperature (LT) phase, the atomic displacements\cite{Fujii} and charge disproportionation occur\cite{Ohama1}, and a spin gap opens\cite{Isobe,Fujii}.
Theoretical studies have proposed that long-range Coulomb repulsion can induce charge ordering in a quarter-filled trellis lattice\cite{Seo,Ohta,Mostovoy}.
In addition, experimental investigations suggest that the transition is not simple; thermal expansion measurements\cite{Koeppen} and resonant x-ray measurements\cite{Nakao} show two-step transitions; NMR measurements suggest the coexistence of the HT and LT phases near $T_{\rm C}$\cite{Ohama1,Ohama2}.
Also, many intermediate phases are observed under pressure by dielectric measurements\cite{Sekine} and x-ray diffraction\cite{Ohwada}.

In the LT phase, superlattice reflections that can be indexed by a face(F)-centered  2$a_p\times$2$b_p\times$4$c_p$ supercell were observed\cite{Fujii}.
X-ray diffraction studies of the LT structure reported the same structure with the space group F$mm2$, which includes three different electronic states of the V sites, V$^{4+}$, V$^{5+}$ and V$^{4.5+}$\cite{Ludecke,Boer}.
This charge distribution is incompatible with $^{51}$V NMR\cite{Ohama1} and resonant x-ray measurements\cite{Nakao}.
Furthermore, $^{23}$Na NMR spectrum measurements show eight independent Na sites, while the space group $Fmm2$ predicts only six Na sites\cite{Ohama2}.
Thus, the LT structure and the charge distribution of $\alpha'$-NaV$_2$O$_5$ have long been controversial.

In this Letter, we report a synchrotron radiation x-ray diffraction study of the  LT structure of $\alpha'$-NaV$_2$O$_5$.
We have found that the structure is monoclinic with the space group $A$112 by refinement on the basis of a domain structure. 
This structure is the only model which is compatible with the resonant x-ray and NMR measurements that have been reported so far.
We have found the charge ordering pattern in the V$_2$O$_5$ plane is of the $zigzag$-type and consists of only V$^{4+}$ and V$^{5+}$ sites.

Single crystals of $\alpha'$-NaV$_2$O$_5$ were grown by the flux method\cite{Isobe2}.
Small high-quality single crystals with typical dimensions 0.05$\times$0.08$\times$0.04 mm$^3$ were used for the x-ray diffraction experiments.
All experiments were carried out at the Photon Factory of KEK.

In order to determine the lattice system below $T_{\rm C}$, high-resolution scattering measurements were performed by using a HUBER four-circle diffractometer with a Si(111) monochromator at BL-4C.
A sample was mounted very carefully to avoid physical stress.
Figure 1 shows the temperature dependence of the peak profiles of (0 2 0) and (6 0 0) reflections.
A peak splitting is observed below $T_{\rm C}$ with (0 2 0) reflection only along the $a^*$-axis, and this splitting disappears above $T_{\rm C}$.
The splitting width increases with decreasing temperature, and is saturated to $\Delta\omega=0.07^\circ$ at 10K.
Since the splittings were not observed in the $(h00)$ and $(00l)$ reflections along the vertical  axes ($i.e.$ (6 0 0) reflection as shown in Fig.1(b)), only  interaxial angle $\gamma$ deviates from 90$^\circ$. 
Thus $\gamma$ value can be directly deduced from the splitting width of the (020) reflections as shown in Fig.1(c).
The LT structure at 11K is an $F$-centered monoclinic system with 2$a_p$=22.65\AA, 2$b_p$=7.179\AA, 4$c_p$=19.030\AA, and $\gamma$=90.035$^\circ$.
Two reflection planes $m_x$ and $m_y$, which exist in the HT phase, disappeared below $T_{\rm C}$.
Thus, the LT structure cannot be described with $Fmm2$ but with $F112$ in the  $2a_p \times 2b_p \times 4c_p$ unit cell.
The conventional unit cell and the symmetry will be described later.
\begin{figure}
\vspace{-2cm}\hspace{-3cm}
\epsfig{figure=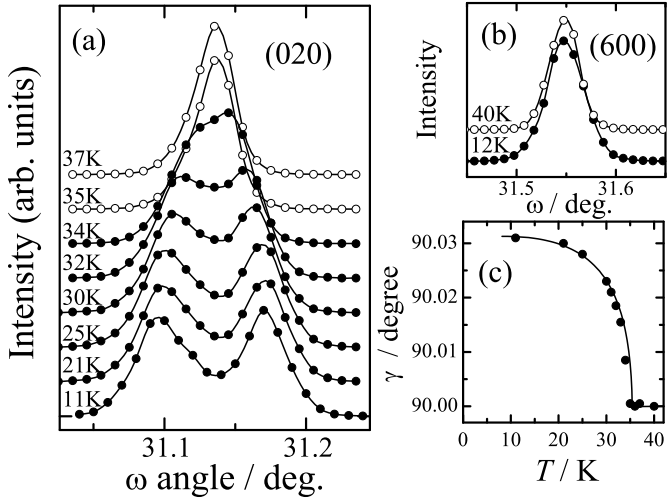,width=12cm}
\caption{Temperature dependence of peak profiles by $\omega$ scan. 
(a) (0 2 0) reflection along the $a^*$ axis. (b) (6 0 0) reflection along the $b^*$ axis.
(c) Temperature dependence of the interaxial angle $\gamma$ angle.}
\end{figure}

For the crystal structure analysis, the reflection data set was measured using the MPD system\cite{Fujiwara} with a two-dimensional (2D) cylindrical imaging-plate detector on BL-1B.
The x-ray wavelength was 0.6888\AA \ calibrated by a CeO$_2$ powder diffraction pattern.
The intensities of the Bragg reflections were measured in a half-sphere of reciprocal space in the range $2\theta <120^\circ$ at 10.0K.
The DENZO program was used for the 2D image processing.
We could not observe the peak splitting of the $(0k0)$ reflections observed in the high-resolution measurements because of the poor resolution of this system.
The intensities of the observed reflections were within the dynamic range of 10$^8$.
The superstructure reflections ($h,k,l=4n+2$) are extremely weak, having an intensity about 10$^{-7}$ times as small as the fundamental intensity.
All superstructure reflections were observed at positions corresponding to the F-centered lattice  $2a_p \times 2b_p \times 4c_p$.
The number of observed unique reflections with $I > 3\sigma (I)$ was 2072.
The Shelx98 program was used for the refinements.

The symmetry of the LT structure must be the lower than orthorhombic according with the Bragg peak splitting.
According to the $F$-lattice extinction rule, the possible space groups of the LT phase are $F112$ and $F$\=1 with the $2a_p \times 2b_p \times 4c_p$ unit cell.
$F11m$ is excluded because it contains the $m_z$ symmetry which is absent in the HT phase.
If the LT structure includes an inversion center, the space group must be triclinic $P$\=1 with $(b_p+c_p)/2 \times (c_p+a_p)/2 \times (a_p+b_p)/2$.
In this structure, the number of inequivalent Na sites is four, a finding that is inconsistent with the $^{23}$Na NMR results\cite{Ohama2}.
In addition, we tried to refine the $P$\=1 models, but obtained only results with a large $R$-factor or with anomalously large residual-peaks.
We therefore concluded that the LT structure has no inversion center.
This conclusion is strongly supported by dielectronic experimental results\cite{Sekine,Poirier}, which indicate a sharp $\lambda$-type peak at $T_{\rm C}$ along the $c$ axis.
Hence the symmetry of the LT phase is indeed monoclinic $F112$, and the conventional crystallographic space group $A112$ with the unit cell $(a_m, b_m, c_m)=(a_p-b_p, 2b_p, 4c_p)$.
This structure has eight inequivalent Na sites and is consistent with the $^{23}$Na NMR results\cite{Ohama2}.

The full refinement with this space group A112 gave almost the same atomic positions as that with F$mm2$\cite{Ludecke,Boer}.
According to the peak-splitting of the (020) reflections shown in Fig.1(a), we can naturally assume that the crystal is a twin structure with a domain $(-a_m, b_m, c_m)=(b_p-a_p, 2b_p, 4c_p)$.
With the full refinement including this twin structure, we obtained a different result from that with the $Fmm2$ ; the $R$-factor was 0.037.
The resulting domain ratio was almost 1:1 (0.498(1):0.502).
Although the difference in the $R$-factors for the $Fmm2$ and $A112$ models is small, the maximum residual peak height 0.69 $e$/\AA$^3$ with $A112$+domain is smaller than 1.17 $e$/\AA$^3$ with $Fmm2$.
It indicates that this refinement is sufficiently reliable.
The final structure parameters are listed in Table I.

\begin{figure}
\epsfig{figure=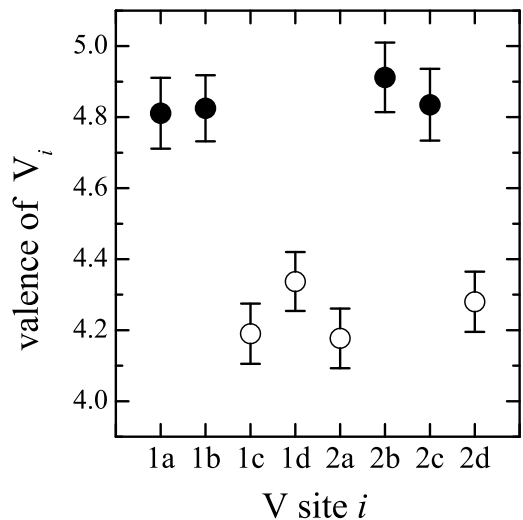,width=8cm}
\caption{The estimated valences of eight V sites by the BVS method.}
\end{figure}

This LT structure consists of eight inequivalent V sites which is constructed in a similar VO$_5$ pyramid structure.
The valence of each V site was calculated by the bond-valence sum (BVS) method\cite{Brese,Brown}.
A bond distance parameter $R$ has been determined for almost any pair of ions in its first coordination shell.
For the V-O pair, $R$ depends on the valence state of V;
$R^{4+}=R$(V$^{4+}-$O) = 1.784\AA \ and $R^{5+}=R$(V$^{5+}-$O) = 1.803\AA\cite{Brese}.
For the HT structure with our refined result at $T$=100K with the space group $Pmmn$, the BVS method gives that a valence of V is +4.56.
Figure 2 shows the calculation results by the BVS method for the eight V sites in the LT phase.
These sites are clearly separated into two groups.
We conclude that the 1c, 1d, 2a, and 2d sites are V$^{4+}$ and the 1a, 1b, 2b, and 2c sites are V$^{5+}$.
This result showing the two separated groups of V sites is in good agreement with the V NMR result.
The deviation of valence from 4+ and 5+ probably reflect the limits of the BVS method.

Figure 3 shows the schematic stacking pattern in the LT phase.
Each ladder consists of a zigzag pattern of V$^{4+}$ and V$^{5+}$ ordering along the $b$ axis alternately as 4+ 5+ 4+ 5+.
This intra-ladder pattern is in good agreement with the theoretical calculations\cite{Seo} and the resonant x-ray scattering\cite{Nakao}.

In a V$_2$O$_5$ plane, the V$^{4+}$ and V$^{5+}$ ions are arranged in a pattern of stripes along the $a_p/3+b_p$(A: $z\sim 1/8,5/8$) or $a_p/3-b_p$(B: $z\sim 3/8,7/8$) directions.
In this charge ordering pattern, the arrangement of V$^{4+}$-magnetic site has a three-fold period along the $a$ axis, which is consistent with the $3a_p^*$ periodically modulated dispersion of magnetic excitation as determined by neutron scattering measurements\cite{Yosihama}.

A distortion from orthorhombic lattice can occur with this diagonal stripe pattern.
However, the A and B stripe patterns are alternately stacked along the $c$ axis, and this stacking is accompanied by only a small deviation of the $\gamma$ angle from 90$^\circ$.
The second nearest two planes have the same stripe direction and out-of-phase stacking.
Hence the four-fold structure along the $c$ axis can be described as the stacking pattern A$^+$B$^+$A$^-$B$^-$(Fig.3).
V ions are arranged as 4+ 4+ 5+ 5+ along the $c$ axis.
This sequence explains why the intensities of the super spots, $l=4n+2$, are extremely weak\cite{NakaoD}.
The LT structure should be stabilized by two factors: the lattice distortion and Coulomb repulsion.
These two different factors may explain the ^^ ^^ devil's staircase''-type of phase transition that occurs under high pressure\cite{Ohwada}.

Another important issue is the origin of the spin singlet ground state and the spin gap in the LT phase.
Two scenarios proposed so far to explain this origin are: (1) an isolated dimer formation of the nearest neighbor (nn) V pairs ({\it e.g.} V$_{\rm 1a}$-V$_{\rm 1b}$) by the inter-ladder exchange interaction\cite{Seo}, and (2) the alternating exchange interaction in the zigzag chain\cite{Mostovoy}.
In the present measurements, we found that the V$^{4+}$-V$^{4+}$ distance in a nn V pair does not change significantly in the HT (3.04\AA\ at 100K) or LT phase (3.02\AA\ at 10K).
On the other hand, even in the LT phase, the magnetic excitations have a strong dispersion along $b^*$ axis\cite{Yosihama}.
These results may support that the dominant origin of the spin singlet is (2) the alternation exchange interaction in a zigzag chain.
\begin{figure}
\epsfig{figure=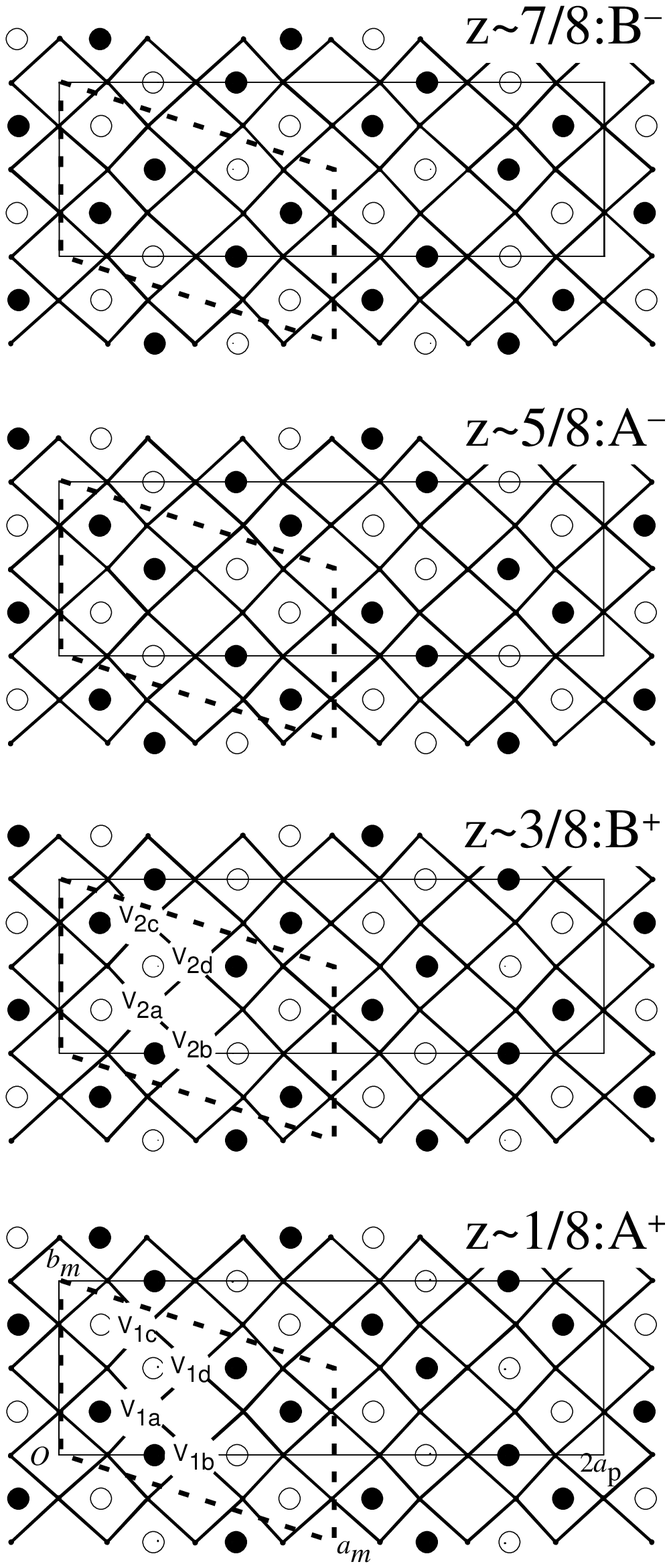,width=7.5cm}
\caption{Stacking pattern along the $c$ axis. 
Projections of the structure in each $ab$ plane are shown.
The unit cells of the supercell with F112 (fine lines) and A2 (gray dashed lines) symmetry are shown.
Solid lines show the basal planes of VO$_5$ pyramids.
The value of $z$ indicate the height of each plane.
Solid and open circles represent V$^{4+}$ and V$^{5+}$ sites, respectively. 
Literature A$^+$,B$^+$,A$^-$,and B$^-$ indicate stacking pattern (see text).
}
\end{figure}

In conclusion, the LT structure of $\alpha'$-NaV$_2$O$_5$ is monoclinic with the space group $A112$ $(a_p-b_p)\times 2b_p \times 4c_p$.
The crystal structure consists of only V$^{4+}$ and V$^{5+}$ sites with a zigzag charge ordering pattern.

We thank Y. Ohta for fruitful discussions.
This work is supported by a Grant-in-Aid for Scientific Research on Priority Areas from the Ministry of Education, Culture, Sports, Science and Technology of Japan.

\narrowtext



\begin{table}
\caption{Atomic coordinations and thermal parameters $B_{eq}$ at 10K with space group A112. 
The temperature factors are refined as anisotropic, but equivalent isotropic ones are listed.
The lattice parameters are $a$=11.879\AA, $b$=7.179\AA, $c$=19.0304\AA, and $\gamma$=107.59$^\circ$. 
\label{table1}}
\begin{tabular}{lcccc}
site & $x$       &    $y$    &     $ z $  &$B_{eq}$ \\
\tableline
V1a  & 0.1501(1) & 0.3254(4) & 0.09938(7)&  0.17(1) \\
V1b  & 0.3497(1) & 0.1751(5) & 0.15079(8)&  0.20(1) \\
V1c  & 0.1542(1) & 0.8272(4) & 0.09572(7)&  0.19(1) \\
V1d  & 0.3457(1) & 0.6734(5) & 0.15403(8)&  0.19(1) \\
V2a  & 0.1543(1) & 0.3276(5) & 0.34641(9)&  0.19(1) \\
V2b  & 0.3504(1) & 0.1753(5) & 0.40048(7)&  0.21(1) \\
V2c  & 0.1501(1) & 0.8255(5) & 0.34887(9)&  0.17(1) \\
V2d  & 0.3449(1) & 0.6728(4) & 0.40437(7)&  0.20(1) \\
Na1a & 0         & 0         & 0.2151(3) &  0.38(5) \\
Na1b & 0.5       & 0         & 0.0367(3) &  0.46(6) \\
Na1c & 0         & 0.5       & 0.2143(3) &  0.42(5) \\
Na1d & 0.5       & 0.5       & 0.0362(3) &  0.55(6) \\
Na2a & 0         & 0         & 0.4650(2) &  0.41(5) \\
Na2b & 0.5       & 0         & 0.2861(3) &  0.47(6) \\
Na2c & 0         & 0.5       & 0.4647(2) &  0.38(5) \\
Na2d & 0.5       & 0.5       & 0.2864(3) &  0.48(6) \\
O11a & 0.0014(4) & 0.2485(8) & 0.1314(1) &  0.37(3) \\
O11b & 0.4992(3) & 0.2515(9) & 0.1228(1) &  0.26(2) \\
O12a &-0.0022(4) & 0.2461(9) & 0.3814(1) &  0.33(3) \\
O12b & 0.4980(3) & 0.2477(9) & 0.3727(1) &  0.21(2) \\
O21a & 0.1780(3) & 0.090(2)  & 0.1273(2) &  0.22(4) \\
O21b & 0.3228(5) & 0.410(2)  & 0.1228(3) &  0.50(6) \\
O21c & 0.1763(4) & 0.588(2)  & 0.1285(2) &  0.25(4) \\
O21d & 0.3250(4) & 0.913(2)  & 0.1208(3) &  0.39(6) \\
O22a & 0.1784(4) & 0.088(2)  & 0.3772(2) &  0.26(5) \\
O22b & 0.3252(5) & 0.410(2)  & 0.3711(2) &  0.45(6) \\
O22c & 0.1758(4) & 0.590(2)  & 0.3782(2) &  0.28(5) \\
O22d & 0.3230(6) & 0.913(2)  & 0.3729(3) &  0.53(6) \\
O31a & 0.1344(4) & 0.316(2)  & 0.0152(3) &  0.51(5) \\
O31b & 0.3627(3) & 0.181(2)  & 0.2358(2) &  0.28(3) \\
O31c & 0.1334(4) & 0.813(1)  & 0.0110(2) &  0.44(5) \\
O31d & 0.3632(3) & 0.687(2)  & 0.2383(2) &  0.31(4) \\
O32a & 0.1336(4) & 0.314(2)  & 0.2617(2) &  0.47(5) \\
O32b & 0.3628(3) & 0.184(2)  & 0.4856(2) &  0.32(4) \\
O32c & 0.1341(4) & 0.812(2)  & 0.2647(3) &  0.48(5) \\
O32d & 0.3630(3) & 0.682(2)  & 0.4884(2) &  0.33(4) \\
\end{tabular}
\end{table}

\end{multicols}

\begin{references}
\bibitem[*]{byline} E-mail address: sawa@science.s.chiba-u.ac.jp.
\bibitem[\dag]{byline2} Present address: Faculty of Science, Tohoku University, Sendai, 980-8577, Japan.
\bibitem{Isobe} M. Isobe and Y. Ueda, J. Phys. Soc. Jpn. {\bf 65}, 1178 (1996).
\bibitem{Ohama1} T. Ohama {\it et al.}, Phys. Rev. B {\bf 59}, 3299 (1999).
\bibitem{Isobe3} M. Isobe and Y. Ueda, J. Alloys Compounds {\bf 262-263},
180 (1997).
\bibitem{Meetsma} A. Meetsma {\it et al.}, Acta Crystallogr.\ Sect.\ C {\bf 54}, 1558 (1998).
\bibitem{Schner} H. G. von Schnering {\it et al.}, Z. Kristallogr. {\bf 213}, 246 (1998).
\bibitem{Smolin} H. Smolinski {\it et al.}, Phys. Rev. Lett. {\bf 80}, 5164 (1998).
\bibitem{Tsuda} K. Tsuda {\it et al.}, J. Phys. Soc. Jpn. {\bf 69}, 1939 (2000).
\bibitem{Horsch} P. Horsch and F. Mack, Eur. Phys. J. B {\bf 5}, 367 (1998).
\bibitem{Fujii} Y. Fujii {\it et al.}, J. Phys. Soc. Jpn. {\bf 66}, 326 (1997). 
\bibitem{Seo} H. Seo and H. Fukuyama, J. Phys. Soc. Jpn. {\bf 67}, 2602 (1998) 
\bibitem{Ohta} S. Nishimoto and Y. Ohta, J. Phys. Soc. Jpn. {\bf 67}, 2996 (1998);P. Thalmeier and P. Fulde, Europhys. Lett. {\bf 44}, 242 (1998).
\bibitem{Mostovoy} M. V. Mostovoy and D. I. Khomskii, Solid State Commun. {\bf 113}, 159 (2000).
\bibitem{Koeppen} M. K\"{o}ppen {\it et al.}, Phys. Rev. B {\bf 57}, 8466 (1998).
\bibitem{Nakao} H. Nakao {\it et al.}, Phys.\  Rev.\  Lett.\  {\bf 85}, 4349 (2000).
\bibitem{Ohama2} T. Ohama {\it et al.}, J. Phys. Soc. Jpn. {\bf 69}, 2751 (2000).
\bibitem{Sekine} Y. Sekine {\it et al.}, to be published in J. Phys. Soc. Jpn. 
\bibitem{Ohwada} K. Ohwada {\it et al.}, Phys. Rev. Lett. {\bf 87}, 086402 (2001).
\bibitem{Ludecke}J. L{\" u}decke {\it et al.}, Phys. Rev. Lett. {\bf 82}, 3633 (1999).; S. van Smaalen and J. L{\" u}decke, Europhys. Lett. {\bf 49}, 250 (2000).
\bibitem{Boer} J. L. de Boer {\it et al.}, Phys.\ Rev.\ Lett.\ {\bf 84}, 3962 (2000).
\bibitem{Isobe2} M. Isobe {\it et al.}, J. Cryst. Growth {\bf 181}, 314 (1997).
\bibitem{Fujiwara} A. Fujiwara et al., J. Appl. Cryst. {\bf 33}, 1241 (2000).
\bibitem{Poirier} M. Poirier {\it et al.}, Phys. Rev. B {\bf 60}, 7341 (1999).
\bibitem{Brese} N. Brese and M. O'Keeffe, Acta. Crystallogr. Sect. B {\bf 47} 192 (1991).
\bibitem{Brown} I. D. Brown, Acta. Crystallogr. Sect. B {\bf 53} 381 (1997).
\bibitem{Yosihama} T. Yosihama {\it et al.}, J. Phys. Soc. Jpn. {\bf 67}, 744 (1998).
\bibitem{NakaoD} H. Nakao, Doctoral thesis, (The University of Tokyo, 1998).
\end{references}
\end{document}